\documentclass[aps,prb,reprint,showpacs,superscriptaddress]{revtex4-2}

\usepackage{graphicx,lipsum}% Include figure files
\usepackage{bm}% bold math
\usepackage[utf8]{inputenc}
\usepackage{amsmath}
\usepackage{float}
\usepackage{amsmath} % or simply amstext
\usepackage{booktabs}
\usepackage{comment}

\usepackage{soul}% To be able to cross out (strike through) with the command \st{strikethrough}.
\usepackage{xcolor}
\definecolor{lightred}{rgb}{1.0,0.3,0.3}
\definecolor{darkred}{rgb}{.8,0.1,0.0}
 
\newcommand\redout{\bgroup\markoverwith{\textcolor{lightred}{\rule[0.5ex]{2pt}{.7pt}}}\ULon}

\begin{document}
\preprint{APS/123-QED}
	
	\title{Crossover from Universal Depinning to Free Domain-Wall Dynamics \\in Ultrathin Iron Garnet Films}

	\author{V. Jeudy}
	\email{vincent.jeudy@universite-paris-saclay.fr}
	\affiliation{Laboratoire de Physique des Solides, Universit\'e Paris-Saclay, CNRS, UMR8502, 91405 Orsay, France.}
	%
	%\author{E. Angeli}
	%\affiliation{Laboratoire de Physique des Solides, Universit\'e Paris-Saclay, CNRS, UMR8502, 91405 Orsay, France.}
	
	\author{D. Gou\'er\'e}
	\affiliation{Laboratoire Albert Fert, CNRS, Thales, Universit\'e Paris-Saclay, 91767, Palaiseau, France.}
	\author{N. Beaulieu}
	\affiliation{LabSTICC,  CNRS,  Universit\'e  de  Bretagne  Occidentale,  29238  Brest, France.}
	\author{S. Husain}
	\affiliation{Laboratoire Albert Fert, CNRS, Thales, Universit\'e Paris-Saclay, 91767, Palaiseau, France.}
	\author{R. D\'iaz Pardo}
	%\affiliation{Laboratoire de Physique des Solides, Universit\'e Paris-Saclay, CNRS, UMR8502, 91405 Orsay, France.}
	\affiliation{Instituto de F\'isica, Universidad Nacional Aut\'onoma de M\'exico, Ciudad de M\'exico 01000, M\'exico.}

	\author{A. Thiaville}
	\affiliation{Laboratoire de Physique des Solides, Universit\'e Paris-Saclay, CNRS, UMR8502, 91405 Orsay, France.}
	\author{J. Sampaio}
	\affiliation{Laboratoire de Physique des Solides, Universit\'e Paris-Saclay, CNRS, UMR8502, 91405 Orsay, France.}
	\author{J-M George}
	\affiliation{Laboratoire Albert Fert, CNRS, Thales, Universit\'e Paris-Saclay, 91767, Palaiseau, France.}
	\author{A. Anane}
	\affiliation{Laboratoire Albert Fert, CNRS, Thales, Universit\'e Paris-Saclay, 91767, Palaiseau, France.}
	\author{J. Ben Youssef}
	\affiliation{LabSTICC,  CNRS,  Universit\'e  de  Bretagne  Occidentale,  29238  Brest, France.}
	
	\date{\today}
	\begin{abstract}
		
	Magnetic domain walls display universal, disorder-controlled elastic dynamics at low drive, and texture-governed free motion at high drive. Here, we establish the crossover mechanism between these two regimes. Using experiments in ultrathin epitaxial iron garnet films and Landau–Lifshitz–Gilbert simulations, including disorder, thermal, and internal texture effects, we uncover a disorder- and temperature-dependent precessional flow that bridges pinned and free dynamics. We further demonstrate that the exceptionally low pinning in garnets arises from the weak coupling between domain walls and disorder, together with a correlation length that exceeds the wall width. 
	
	\end{abstract}
	
	% insert suggested PACS numbers in braces on next line
	%75.78.Fg: Dynamics of magnetic domain structures, 
	%68.35.Rh: Phase transitions and critical phenomena,
	% 05.70.Ln: Nonequilibrium and irreversible thermodynamics, 
	% 64.60.Ht: Dynamic critical phenomena, 
	%47.54.−r: Pattern selection; pattern formation
	%
	
	\maketitle
	 Understanding the dynamics of domain walls (DWs) in ultrathin films, as well as the role of intrinsic material defects that introduce stochasticity, is central to the development of emerging spintronic devices~\cite{velez_natcom_2019,caretta_apl_mat_2024}. It is also of great interest for fundamental physics, since DWs provide a model system for studying driven elastic objects in disordered media~\cite{chauve_2000,Ferre_CRP_2013_review}.
	 Experimentally, identifying materials with weak defect landscapes is crucial~\cite{De_Leeuw_RepProgPhys_1980,malozemoff,Wei_NatMat_2022}, as this enables access to a wide range of pinned and free dynamical regimes. By contrast, in systems with a large depinning threshold—such as Pt/Co/Pt~\cite{jeudy_PRB_2018_DW_pinning}—pinning effects dominate and mask most free-flow dynamics.
	 In this context, single-crystalline epitaxial iron garnets are particularly attractive, since the literature reports exceptionally low depinning thresholds~\cite{De_Leeuw_RepProgPhys_1980,malozemoff}.
	 Their study has gained renewed interest with the development of ultrathin films, where interface engineering enables spin-orbit torque (SOT) driven motion of magnetic textures at low current densities and with high mobilities~\cite{velez_natcom_2019,caretta_apl_mat_2024}.
	 Yet, pinned and high-drive regimes remain scarcely explored, and the microscopic origin of weak pinning is still unclear. Moreover, ultrathin films, with suppressed thickness non-uniformities of DW-texture, offer unique opportunities to systematically address the crossover between pinned and free DW motion, which remains an open issue.

	%In materials with a large depinning threshold ($f_d \gg f_w$), such as Pt/Co/Pt~\cite{jeudy_PRB_2018_DW_pinning}, the pinning masks most of the flow regimes, so that only the high-drive precessional asymptotic regime~\cite{Schryer_JAP_1974,malozemoff} may be observed. The other flow regimes are accessible in films and multilayers with reduced $f_d$-values~\cite{dourlat_prb_2008,Beach_NatMat_2005,yamada_apex_2011} and/or with high $f_w$-values due to the Dzyaloshinskii-Moriya interaction~\cite{thiaville_EPL_2012}.
 	%\VJ{For garnet films, . The particularly low reported depinning threshold, as well as 

	On the theoretical side, rather different models are used to describe the pinned and free dynamics. The free dynamics is commonly derived from the Landau-Lifshitz-Gilbert (LLG) equation. DW motion relies on the coupling between the drive and the DW internal magnetic texture~\cite{Schryer_JAP_1974,malozemoff,thiaville_EPL_2012}. The description of complex DW behaviors involving non-homogeneous DW magnetic textures requires numerical micromagnetic simulations~\cite{yamada_apex_2011,herranen_PRL_2019}.
	Some attempts have been made to reproduce experimentally measured depinning thresholds by including disordered micromagnetic properties in simulations~\cite{jue_PRB_2016, moretti_PRB_2017, herranen_PRL_2019}. However, key ingredients for the description of pinned regimes are not taken into account, e. g. the thermal activation, or they are adjusted purely empirically, like the amplitude and correlation length of the disorder.
	In contrast, pinned domain wall (DW) dynamics is largely independent of the DW internal texture~\cite{lemerle_PRL_1998_domainwall_creep, albornoz_PRB_2024}. The universal behaviors known as the thermally activated creep regime and depinning transition are well captured by the quenched Edwards-Wilkinson (qEW) minimal model~\cite{jeudy_PRL_2016_energy_barrier,diaz_PRB_2017_depinning}, which describes a DW as an elastic line placed in a short-range random pinning landscape, submitted to thermal activation~\cite{kolton_prb_2009_pathways}. 
	Recent theoretical developments based on scalar field models~\cite{caballero_PRE_2018} and simulations based on the LLG equation~\cite{guruciaga_JStatMech_2021} have tried to go beyond the qEW minimal model. However, the presumed fixed internal DW magnetic texture leaves aside the complex free DW dynamics and the crossover from the pinned regimes. 
	Therefore, to gain deeper insight into DW dynamics, it is essential to develop simulations that incorporate disorder, thermal fluctuations, and internal magnetic texture effects, thereby bridging the pinned and free DW motion within a unified framework.
	
	%Therefore, to better understand DW dynamics, it would be particularly interesting to develop simulations accounting for disorder, thermal, and internal DW magnetic texture effects, thus allowing one to bridge the pinned and free DW motion in a single model.
	
	In this Letter, we extract the correlation length of the DW–disorder interaction from the analysis of DW dynamics and develop LLG-based simulations including disorder and thermal effects. The excellent agreement with experiments across a magnetic field range spanning two orders of magnitude reveals the nature of the crossover between pinned and free DW motion.
	% The excellent agreement with experiments over two decades in field uncovers the nature of the crossover between pinned and free DW motion.
			
\begin{figure*}	
	\includegraphics[scale=0.57]{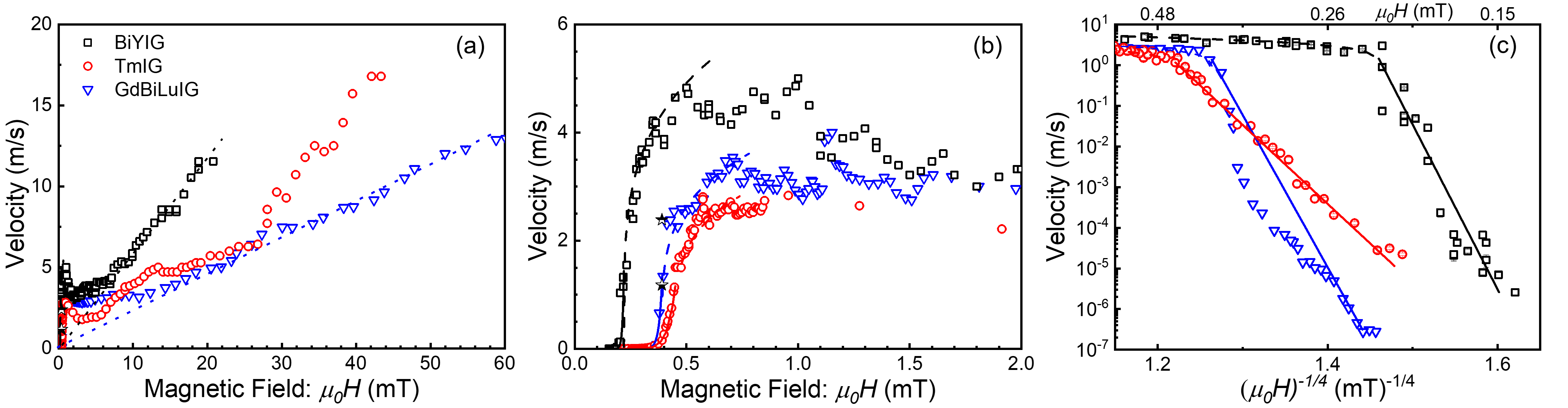}
	\caption{\textbf{Domain wall dynamics in garnet films}. (a) DW velocity versus magnetic field ($\mu_0 H$) showing the flow regimes. The short-dot lines reflect the asymtotic precessional flow regime. (b) Zoom of the same curves highlighting the depinning regime and the crossover to the flow regime. (c) Same curves presented in log scale versus $(\mu_0 H)^{-1/4}$ revealing the creep regime.
	In (b) and (c), the solid and dash lines represent the predictions of Eqs.~\ref{eq: 1} for the creep and depinning, respectively.} 
	\label{fig:1}
\end{figure*} 
 
	\textit{Experimental techniques.} 
	% Samples
	%The experiments were performed on iron garnet epitaxial films grown along the [111] direction by different deposition methods. 
	The experiments were performed on epitaxial iron garnet films grown along the [111] direction, exhibiting an out-of-plane easy magnetization axis, using different deposition methods.
	Here, we compare typical DW dynamics obtained for:
	% TCM19 : (GdBiLu)3(FeAl)5O12/GGG
	(1) a 44~$nm$ thick film of (GdBiLu)$_{3}$(FeAl)$_{5}$O$_{12}$ (GdBiLuIG) deposited by liquid phase exitaxy (LPE)~\cite{benyoussef_thesis_1989} on a Gadolinium Gallium Garnet (GGG) substrate,
	% Al5450
	(2) a  16~nm-thick Bi$_{0.7}$Y$_{2.3}$Fe$_{5}$O$_{12}$ (BiYIG) garnet film obtained by pulsed laser deposition (PLD)~\cite{Soumah_natcom_2018} on a substituted GGG substrate,
	% B2A5 Pt/Tm3Fe5O12/GGG
	(3) a 15~nm-thick Tm$_{3}$Fe$_{5}$O$_{12}$/GGG (TmIG) garnet film  grown by off-axis sputtering~\cite{husain_nanolett_2024} and covered by 6~nm of Pt. 
	The micromagnetic parameters of the three samples are reported in Table I.
	% Domain wall dynamics
	The DW motion was driven by perpendicular magnetic field pulses of adjustable amplitude ($\mu_0H=0-65~\mathrm{mT}$), which were produced by a 75 turns coil (diameter $\approx$ 1 mm, rise and fall time $\tau \approx 0.2~\mathrm{\mu s}$) mounted onto the films.
	The DW displacement was observed using a magneto-optical Kerr microscope (resolution $\sim$ 1~$~\mathrm{\mu m}$). The velocity $v$ is defined as the ratio between average displacement $\Delta x$ and pulse duration $\Delta t$. $\Delta t$ was adjusted between 1~$~\mathrm{\mu s}$ and 120~s to measure a significant displacement ($\Delta x >10~\mathrm{\mu m}$), according to each value of $v(\mu_0H)$.

	\textit{Numerical simulation.}
	For the  simulation, the Landau-Lifshitz-Gilbert (LLG) equation is solved using MuMax3, a GPU-based micromagnetic finite-difference solver~\cite{vansteenkiste_AIP_2014}.
	The	dimensions of the reference frame in the $x, y, z$ directions are $L_x L_y L_z = 2.56~\mathrm{\mu m} \times 10.24~\mathrm{\mu m} \times 15~\mathrm{nm}$, corresponding to $512 \times 2048 \times 1$ cells. Periodic boundary conditions are used in the $y$ direction.
	An initially flat DW was placed parallel to the $y-z$ plane separating two regions with opposite magnetization ($M_z=M_s$ for $x<L_x/2$ and $M_z=-M_s$ for $x>L_x/2$). 
	The magnetic field was applied along the $z$-direction to move the DW in the $x>0$ direction. The reference frame was translated along the $x$-direction during the motion in order to maintain the DW average position close to $L_x/2$. 
	The micromagnetic parameters are those of the TmIG film (see Table~\ref{table:table1}, with the gyromagnetic factor $g = 1.63$), which presents the richest DW dynamics. The damping constant ($\alpha= 0.02$) was chosen so to best reproduce the experimentally observed flow regime. The temperature was set to %$T=0~\mathrm{K}$ and 
	$T=293~\mathrm{K}$. 
	To emulate the disorder, we introduced a Voronoi tessellation with a cell size parameter $b=300~\mathrm{nm}$ close to the values of the DW-disorder interaction length scale $\xi$ deduced from experiments (see the paragraphs just after and Table~\ref{table:table1}). The values of  $K_u$ followed a normal distribution and were randomly assigned to the Voronoi cells. The width of the distribution was adjusted ($\delta K_u/K_u= 4.2~\mathrm{\%}$) to obtain the best agreement of the DW velocity with experimental results for the creep and depinning regime. 
	%	$\delta \sigma t \approx \delta K_u \Delta t$ with $ \sigma=4\sqrt{A (K_u -\mu_0 M_s^2/2)}$ $\delta \sigma t=27 f N$ for $t=15 nm$.
	% 
	%
	Statistical variations were obtained by using different disorder maps or a single disorder map with different exchange coupling between cells.
	% To ensure that the DW reached a steady-state motion, the simulations were run for duration up to $30 ~\mathrm{\mu s}$, which is more than an order of magnitude longer than those typically reported in the literature
	To ensure steady-state DW motion, simulations were run up to $30~\mathrm{\mu s}$, more than one order of magnitude longer than typical literature values~\cite{van_de_wiele_PRB_2012,moretti_PRB_2017,herranen_PRL_2019} (see supplemental material for more details on the simulation).

	\textit{Qualitative description of the experimental velocity curves.}
	The velocity curves obtained for the three samples (see Fig. \ref{fig:1}) present similar shapes and highlight the different DW free and pinned dynamical regimes. At the lowest drive (see Fig.~\ref{fig:1} c), the velocity follows rather well the creep law ($\ln v \sim H^{-1/4}$).
	The weak discrepancy for the GdBiLuIG film is most probably due to spatial inhomogeneities of the disorder, which limits investigations to short scale ($10 \mu m$) DW displacements. A steep slope (see Fig.~\ref{fig:1} b) leading to a maximum of the velocity  reveals the end of the creep regime and the depinning transition.
	Flow regimes (see Fig.~\ref{fig:1} a) start after the peak with a short plateau followed by a linear regime corresponding to the expected asymptotic precessional flow. For the TmIG film, the slope increase starting close to $\mu_0 H_d=30~\mathrm{mT}$ is associated with a DW instability, as discussed later.  
	
		\begin{table*}[h!]
		\centering
		\begin{ruledtabular}
			%\begin{tabular}{|p{5cm}|c|c|c|r|r|}
			\begin{tabular}{|l|c|c|c|c|c|c|c|c|c|c|c|c|}
				
				Sample & \bf{$t$} &\bf{$ \mu_0M_s$} & \bf{$ K_{eff}$} & \bf{$ A$} & \bf{$\Delta$} &\bf{$\sigma$} &  \bf{$ \mu_0H_d$} & \bf{$v(H_d)$} & \bf{$ T_d$} & \bf{$\xi$} & \bf{$f_{pin}$} & \bf{$\alpha$} \\
				
				\hline
				& \bf{$nm$} &\bf{$ mT$} & \bf{$kJ/m^3$} & \bf{$ pJ/m$} & \bf{$ nm$} &\bf{$\mu J/m^2$} &  \bf{$mT$} & \bf{$m/s$} & \bf{$   10^3 K$} & \bf{$ nm$} & \bf{$fN$} &\\
				
				\hline
				GBLIG 
				& 44(2) & 120(10) & 5.25(0.5) & 3.5 & 26(2)& 540(30)&
				0.39(0.05)& 1.18 (0.05)&32(4)&210(25)  &13(2) & 0.049(0.003)  \\
				BiYIG
				&  16(1) & 180(20) & 10.4(2.5) & 3.5 & 18(2)& 760(90)&
				0.22(0.01)& 1.44(0.05)&40(4)&470(60)&3.5(0.4)& 0.18(0.01) \\			
				TmIG
				&  15(1) & 140(10) & 3.3(0.3) & 3.26(0.01) & 32(2)& 430(20)&
				0.45(0.02)& 1.16(0.05)&16(2)&270(30) & 4.6(0.5)& 0.02(0.01)\\
				
				\hline
				Simul
				& 15 & 140 & 3.3 & 3.5 & 32 & 434 & 0.465(0.050)& 1.2 (0.2)& 16(4) & - & - & 0.02\\
				 %$270 (60)$
			\end{tabular}
			\caption{\label{table:table1}  \textbf{Micromagnetic and depinning parameters.}
				For each sample, the table indicates the thickness $t$, the magnetization saturation $\mu_0 M_s$ deduced from squid measurement. The effective anisotropy constant  $K_{eff}=\mu_0 M_s H_{eff}/2$ is obtained from anisotropy field measurements, the exchange constant $A$ was taken from Refs.~\cite{benyoussef_thesis_1989,Soumah_natcom_2018} for GBLIG and BiYIG, and extracted from Brillouin light scattering measurements for TmIG.  The domain wall thickness parameter and energy are calculated from $\Delta=\sqrt{A/K_{eff}}$ and $\sigma=4\sqrt{A K_{eff}}$, respectively.  
				The depinning temperature $T_d$ and magnetic field $H_d$ and velocity at the depinning $v(H_d)$ are deduced from simultaneous fits of Eqs.~\ref{eq: 1} to experimental creep and depinning velocities.
				The length scale $\xi$ and force $f_{pin}$ of the DW-disorder interaction are calculated as explained in the text~\cite{gehanne_PRR_2020}.
				%from  $\xi \sim [(k_B T_d)^2/(2 \mu_0 M_s H_d \sigma t^2)]^{1/3}$, and 
				%$f_{p} \sim \sqrt{2 \mu_0 H_d M_s t k_B T_d}$, respectively. 
				%For the GBLIG and BiYIG films, the damping factor $\alpha$ was deduced from the prediction~\cite{malozemoff} $v=\gamma \Delta \alpha/ (1+\alpha^2)$ for the precessional asymptotic regime.
				%For the TmIG films, $\alpha$ was deduced from the comparison with the simulation. 
				The damping factor $\alpha$ was determined from the precessional asymptotic relation $v = \gamma \Delta \alpha / (1 + \alpha^2)$~\cite{malozemoff} for GBLIG and BiYIG, and from simulation comparison for TmIG.
				The numbers between parentheses are the error bars. 
				%{\JS it's simpler to use $1.44\pm0.05$} but it takes more space and overflows. 
			} 
		\end{ruledtabular}
	\end{table*}

	\textit{Pinning-dependent regimes and DW-disorder interaction.}
	 In order to analyze DW dynamics quantitatively, we first address the pinning dependent regimes and the DW-disorder interaction. 
	 %The whole set of dynamical regimes will then be discussed on the basis of a numerical simulation. 
	 In the creep and depinning regimes, the DW velocity can be described self-consistently by the following relations~\cite{diaz_PRB_2017_depinning,jeudy_PRB_2018_DW_pinning}:
	\begin{equation}
		v(H)=\left \{
		\begin{array}{lr}
			v(H_d)\exp (-\frac{\Delta E(H)}{k_B T}) & {\rm creep:} H<H_d\\
			\frac{v(H_d)}{x_0}(\frac{T_d}{T})^{\psi}(\frac{H-H_d}{H_d})^\beta & {\rm depinning:} H \gtrsim H_d, \\
		\end{array}
		\right.
		\label{eq: 1}
	\end{equation}
	where $\Delta E(H)=k_B T_d((H/H_d)^{-\mu}-1)$ is the energy barrier of creep regime, $H_d$ the depinning field, $k_B T_d$ the characteristic height of effective pinning barrier, and $v(H_d)$ the velocity at depinning threshold ($\Delta E \rightarrow 0$). In Eqs.~\ref{eq: 1}, $\mu=1/4$, $\beta=0.25$, and  $\psi=0.15$ are universal critical exponents and $x_0=0.65$ a universal constant~\cite{diaz_PRB_2017_depinning,jeudy_PRB_2018_DW_pinning}. The non-universal parameters ($H_d,v(H_d)$, and $T_d$) characterize DW depinning for each material.
	%
	%{\JS [perhaps new paragraph here.] [if I understand correctly, this sentence should come after the next one which explains how the parameters were found]} The rather good agreement between the experiments and predictions of Eqs.~\ref{eq: 1} (see Figs.~\ref{fig:1} b and c) confirms {\JS the} a compatibility with the qEW universality class. 
	The depinning parameters were deduced from a simultaneous fit of Eqs.~\ref{eq: 1} for each film. 
	%The obtained values are reported in Table~\ref{table:table1}. 
	The agreement between the experiments and predictions (see Figs.~\ref{fig:1} b and c) confirms the compatibility with the qEW universality class. 
	%{\JS [*** say here that the model reproduces the observations, after explaining how the Td Hd were found.]} 
	The values of the depinning parameters (see Table~\ref{table:table1}) are rather close for the three films, suggesting a weak dependency of DW pinned dynamics on the synthesis method and on the precise composition of the iron garnet films.
	The depinning field $\mu_0 H_d=0.22-0.45~\mathrm{mT}$ is about one to two orders of magnitude lower than that of typical CoFeB and Pt/Co/Pt films~\cite{jeudy_PRB_2018_DW_pinning}, respectively,  
	%\JSremove{close to one and two order of magnitude lower than for typical CoFeB and Pt/Co/Pt films~\cite{jeudy_PRB_2018_DW_pinning}, respectively,} {\JS that of CoFeB and about two orders of magnitude lower than that of Pt/Co/Pt films~\cite{jeudy_PRB_2018_DW_pinning},}
	%
	thus allowing the observation of free DW dynamics over a wide range of magnetic field. 
	%
	%The depinning field $\mu_0 H_d=0.22-0.45mT$ is close to \JSremove{ one and two order of magnitude lower than for typical CoFeB and Pt/Co/Pt films~\cite{jeudy_PRB_2018_DW_pinning}, respectively,} {\JS that of CoFeB and about two orders of magnitude lower than that of Pt/Co/Pt films~\cite{jeudy_PRB_2018_DW_pinning},} thus allowing the observation of free DW dynamics over a wide range of magnetic field. 
	%
	%(The steep slope at depinning threshold is due~\cite{diaz_PRB_2017_depinning} to a high ratio between thermal activation energy ($k_B T$) and effective pinning barrier ($T_d/T\approx 110-140 $).)
	
	The DW-disorder interaction is investigated using the scaling model developed in Ref.~\cite{gehanne_PRR_2020}. The DW-disorder interaction length scales as $\xi \sim [(k_B T_d)^2/(2 \mu_0 M_s H_d \sigma t^2)]^{1/3}$.
	%
	%the characteristic length (Larkin) of collective pinning $L_c \sim [(\sigma k_B T_d)/(4M_s^2 t H_d^2)]^{1/3}$, and the strength of pinning disorder $D^2 \sim (4\pi)^{1/6}[(k_BT_d)^5]/[\sigma t L_c]$. 
	%
	The obtained values (see Table~\ref{table:table1}), $\xi$ ($=200-260~\mathrm{nm}$) are much larger than the DW width parameter ($\xi/\Delta = 16-35 \gg1 $). This suggests that DW pinning in iron garnets is associated to a weak concentration of pinning sites 
	%($n=1/b^2$) 
	with a disorder correlation length $b~\approx \xi$ significantly larger than $\Delta$~\cite{nattermann_prb_1990}. This is starkly different from metallic ultra-thin ferromagnets, which present significantly higher concentrations of defects ($\xi/\Delta = 1-3$)~\cite{jeudy_PRB_2018_DW_pinning}, and a lower correlation length of the disorder ($b<\Delta~\approx \xi$)~\cite{gehanne_PRR_2020}.
	Moreover, we can compare the ratio between the DW line energy $\sigma t$ and the force of DW-disorder interaction~\cite{gehanne_PRR_2020} $f_{p} \sim (b/\xi)\sqrt{2 \mu_0 H_d M_s t k_B T_d}$ (see Table~\ref{table:table1}). The ratio $\sigma t/f_{p} = 1400-3400$ is two orders of magnitude larger than for Pt/Co/Pt and Pt/Co/AlOx ultra-thin films~\cite{gehanne_PRR_2020} ($\sigma t/f_{p} \approx 10$, $t \approx 1~\mathrm{nm}$)  reflecting the much lower DW-disorder interaction in iron garnets. 

	\begin{figure*}	
		\includegraphics[scale=0.57]{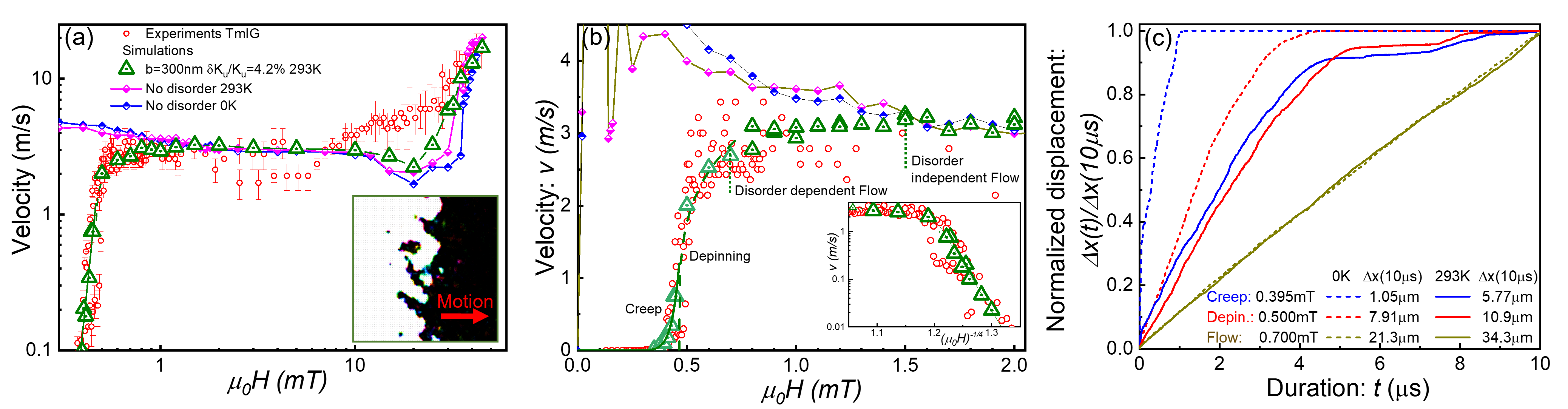}
		\caption{\textbf{Comparison between numerical simulations and experimental results for the TmIG film}. (a) DW velocity versus $\mu_0 H$ in log-log scale, showing the full range of DW dynamics. Simulations were performed for $T=293~\mathrm{K}$ with (green triangles) and without disorder (half-filled down magenta diamonds) and for $T=0~\mathrm{K}$ without disorder (half filled up violet diamonds). The green solid and dash lines correspond to predictions of Eqs.~\ref{eq: 1}, for the creep and depinning regime, respectively. Inset: Propagation front within the instability domain observed at $\mu_0H=40.0~\mathrm{mT}$ with closed areas located in front and behind. The image size is $L_x L_y = 5.12 \times 5.12 ~\mathrm{\mu m^2}$. (b)  Close-up of the velocity curves highlighting the crossover between pinned and free DW dynamics. 
		%\VJ{the disorder and temperature-dependent flow regime the creep, corresponding to the crossover between depinning flow regimes
		Inset: Same curve as presented in log scale versus $(\mu_0 H)^{-1/4}$ to emphasize the creep regime. (c) Simulated DW displacement $\Delta x(t)$ versus time for $T=0~\mathrm{K}$ and $T=293~\mathrm{K}$, and under different magnetic fields ($\mu_0H=0.395$, $0.500$, and $0.700~\mathrm{mT}$), corresponding to the creep, depinning, and flow regimes. Displacements are normalized to $\Delta x(t=10~\mathrm{\mu s}, \mu_0H)$, with the normalization values indicated at the bottom of the panel. 
		} 
		\label{fig:2}
	\end{figure*}
	
	{\it Free DW dynamics.} We now extend our analysis to the whole DW dynamics through a detailed comparison between experiments and micromagnetic simulations.
	As shown in Figs.~\ref{fig:2}a and b, the simulation reproduces rather well the crossover between the pinned and flow dynamics and (for $\mu_0H>2~\mathrm{mT}$) captures the key trends of the free dynamics observed in the experiments.
	To differentiate between disorder-dependent and disorder-independent regimes, and to analyze the role of thermal effects, we performed additional simulations without disorder for $T=0$ and $293~\mathrm{K}$. 
	As expected at low drive, the DW velocities are higher without than with disorder.  Above $\mu_0H \approx 1.5~\mathrm{mT}$, the curves remain superimposed until $\mu_0H \approx 10~\mathrm{mT}$, thus reflecting a flow behavior independent of both disorder and temperature.
	The damping parameter was obtained by adjusting the simulation to match the experimental results in the range $1.5 mT< \mu_0H <10~\mathrm{mT}$, and yielded a value of $\alpha = 0.02$. Comparable values of $\alpha$ were also obtained from the slope of the asymptotic precessional regime (see Fig.\ref{fig:1}a), as reported in Table 1 for both the GdBiLuIG and BiYIG films. These values are significantly higher than those extracted from ferromagnetic resonance (FMR) measurements ($7 \times 10^{-3}$ for GdBiLuIG and $0.7$–$1.9 \times 10^{-3}$ for BiYIG), as encountered in other materials~\cite{dourlat_prb_2008,chauleau_PRL_2014}, and which may be
	due to non-local damping effects~\cite{chauleau_PRL_2014}.
	%, and which may be due to non-local damping effects~\cite{chauleau_PRL_2014}.
	
	Let us discuss the nature of the flow regimes observed at low and high drive. The curves obtained with no disorder present the expected Walker peak for $\mu_0H_W=0.125~\mathrm{mT}$ (and $v_W=21~\mathrm{m/s}$, out of range in Figs.~\ref{fig:2}a and b). Since $H_W<H_d$, the experimentally observed flow regimes are precessional. 
	Moreover, a strong mobility increase (see Fig.~\ref{fig:2} a) is observed for $\mu_0H>10~\mathrm{mT}$ (and is favored by thermal activation and disorder). Its origin is highlighted in the inset of Fig.\ref{fig:2} a, which reveals an instability of the DW with domains of opposite magnetization orientations on either side of the DW propagation front. A deeper understanding of this interesting instability is beyond the scope of our present study.
	
	{\it Crossover regime between pinned and free dynamics.} As simulations take into account the DW internal texture, disorder, and temperature, it is particularly stimulating to investigate the crossover between pinned and free dynamics observed in the range $\approx 0.700~\mathrm{mT} <\mu_0H< 1.5~\mathrm{mT}$ (see Fig.~\ref{fig:2} b). 
	Indeed, this dynamical regime cannot be described by the universal qEW elastic line model for the creep and depinning regimes nor by a disorder-independent free DW motion derived from the LLG equation. Note that the crossover regime is most probably a ubiquitous phenomenon of DW dynamics since it has also been observed for Pt/Co/Pt films with depinning field lying in the asymptotic precessional flow ($H_d \gg H_w$)~\cite{diaz_PRB_2017_depinning}, while it is encountered close to $H_w$ in garnet films. 
	
	To address the pinned or free nature of DW dynamics, the simulated time response of the initially flat DW to a magnetic field step
	%, for $T=0\mathrm{K}$ and $T=293\mathrm{K}$, 
	proved to be particularly fruitful.  
	Fig.~\ref{fig:2}c compares the average DW displacement $\Delta x(t)$ in the creep and depinning regimes, and just at the beginning of the crossover.
	For $T=293\mathrm{K}$, $\Delta x(t)$ in the creep and depinning presents two regimes with a steep slope ($t<4~\mathrm{\mu s}$) followed by a strongly reduced slope (by a factor $\approx 20$). As the slope of the second regime remains constant 
	%for the longest simulation duration ($\approx 30~\mathrm{\mu s}$) 
	(see supplemental material), it corresponds to a steady DW motion, while the first regime corresponds to a transient. 
	For $T=0\mathrm{K}$, the steady regimes fully disappear. This collapse of the steady regime reflects its thermally activated nature for the creep motion~\cite{jeudy_PRL_2016_energy_barrier}. For the depinning, it is associated with the increase of the depinning field ($\mu_0H_d(293\mathrm{K})=0.45 \pm 0.05~\mathrm{mT}$ and $\mu_0H_d(0\mathrm{K})=0.63 \pm 0.02~\mathrm{mT}$) as the temperature decreases~\cite{diaz_PRB_2017_depinning}.
	For the transient regime, the significant reduction of the displacement duration between $293\mathrm{K}$, and $0\mathrm{K}$ reveals finite temperature effects. However, its persistence without thermal activation suggests another mechanism already discussed from numerical studies of the non-steady motion of an elastic line in two-dimensional random disorder in Ref.~\cite{ferrero_PRE_2013} performed at $0\mathrm{K}$. Below the depinning threshold, an initially flat line undergoes quasi-free motion and becomes increasingly rough as interactions with the disorder strengthen, leading to a gradual reduction in velocity until it eventually stops.
	Therefore, our simulation evidences the transient roughening transition of the DW and the steady creep and depinning motion at finite temperature.
	
	In contrast, for $\mu_0H=0.700~\mathrm{mT}$, $\Delta x$ presents a single linear variation with $t$. 
	The absence of transient regime indicates that the DW is no longer pinned by the disorder, as expected for a flow regime. In this particular flow regime, the disorder reduces the DW velocity compared to a fully free DW (see Fig.~\ref{fig:2} b) and finite temperature effects enhance DW displacement (see Fig.~\ref{fig:2} c).  
	Thus, we reveal the nature of the crossover dynamics between pinned and free DW regimes, which corresponds to a strongly disorder- and temperature-dependent flow regime.

	{\it Conclusion.} 
	%We have demonstrated that DW motion in ultra-low pinning iron garnet films can be quantitatively simulated from the Landau–Lifshitz–Gilbert framework, including disorder and thermal effects. Our results clarify the crossover between pinned and free dynamics and pave the way for the study of other magnetic textures such as skyrmions. For disordered elastic systems, our finding should help to better understand the transition from pinned to sliding interface motion with increasing drive, which is a fundamental nonequilibrium process.
	%
	We have shown that domain-wall motion in ultralow-pinning iron garnet films can be quantitatively reproduced within the Landau–Lifshitz–Gilbert model  when disorder and thermal effects are included. This unified approach captures the full crossover from the universal depinning regime to free flow, bridging the gap between pinned and unpinned dynamics. Beyond domain walls, our findings provide a general framework for understanding driven disordered interfaces and open new perspectives for the study of other magnetic textures, such as skyrmions.
	
	%\newpage
	\begin{acknowledgments}
		This work has been supported by the French National Research Agency under the project ’DeMIuRGe’ ANR-22-CE30-0014.
		
	\end{acknowledgments}

%\sloppy	
\bibliography{refs_Walker,refs_simul}

%apsrev4-2.bst 2019-01-14 (MD) hand-edited version of apsrev4-1.bst
%Control: key (0)
%Control: author (8) initials jnrlst
%Control: editor formatted (1) identically to author
%Control: production of article title (0) allowed
%Control: page (0) single
%Control: year (1) truncated
%Control: production of eprint (0) enabled
\begin{thebibliography}{31}%
\makeatletter
\providecommand \@ifxundefined [1]{%
 \@ifx{#1\undefined}
}%
\providecommand \@ifnum [1]{%
 \ifnum #1\expandafter \@firstoftwo
 \else \expandafter \@secondoftwo
 \fi
}%
\providecommand \@ifx [1]{%
 \ifx #1\expandafter \@firstoftwo
 \else \expandafter \@secondoftwo
 \fi
}%
\providecommand \natexlab [1]{#1}%
\providecommand \enquote  [1]{``#1''}%
\providecommand \bibnamefont  [1]{#1}%
\providecommand \bibfnamefont [1]{#1}%
\providecommand \citenamefont [1]{#1}%
\providecommand \href@noop [0]{\@secondoftwo}%
\providecommand \href [0]{\begingroup \@sanitize@url \@href}%
\providecommand \@href[1]{\@@startlink{#1}\@@href}%
\providecommand \@@href[1]{\endgroup#1\@@endlink}%
\providecommand \@sanitize@url [0]{\catcode `\\12\catcode `\$12\catcode
  `\&12\catcode `\#12\catcode `\^12\catcode `\_12\catcode `\%12\relax}%
\providecommand \@@startlink[1]{}%
\providecommand \@@endlink[0]{}%
\providecommand \url  [0]{\begingroup\@sanitize@url \@url }%
\providecommand \@url [1]{\endgroup\@href {#1}{\urlprefix }}%
\providecommand \urlprefix  [0]{URL }%
\providecommand \Eprint [0]{\href }%
\providecommand \doibase [0]{https://doi.org/}%
\providecommand \selectlanguage [0]{\@gobble}%
\providecommand \bibinfo  [0]{\@secondoftwo}%
\providecommand \bibfield  [0]{\@secondoftwo}%
\providecommand \translation [1]{[#1]}%
\providecommand \BibitemOpen [0]{}%
\providecommand \bibitemStop [0]{}%
\providecommand \bibitemNoStop [0]{.\EOS\space}%
\providecommand \EOS [0]{\spacefactor3000\relax}%
\providecommand \BibitemShut  [1]{\csname bibitem#1\endcsname}%
\let\auto@bib@innerbib\@empty
%</preamble>
\bibitem [{\citenamefont {V{\'e}lez}\ \emph {et~al.}(2019)\citenamefont
  {V{\'e}lez}, \citenamefont {Schaab}, \citenamefont {W{\"o}rnle},
  \citenamefont {M{\"u}ller}, \citenamefont {Gradauskaite}, \citenamefont
  {Welter}, \citenamefont {Gutgsell}, \citenamefont {Nistor}, \citenamefont
  {Degen}, \citenamefont {Trassin}, \citenamefont {Fiebig},\ and\ \citenamefont
  {Gambardella}}]{velez_natcom_2019}%
  \BibitemOpen
  \bibfield  {author} {\bibinfo {author} {\bibfnamefont {S.}~\bibnamefont
  {V{\'e}lez}}, \bibinfo {author} {\bibfnamefont {J.}~\bibnamefont {Schaab}},
  \bibinfo {author} {\bibfnamefont {M.~S.}\ \bibnamefont {W{\"o}rnle}},
  \bibinfo {author} {\bibfnamefont {M.}~\bibnamefont {M{\"u}ller}}, \bibinfo
  {author} {\bibfnamefont {E.}~\bibnamefont {Gradauskaite}}, \bibinfo {author}
  {\bibfnamefont {P.}~\bibnamefont {Welter}}, \bibinfo {author} {\bibfnamefont
  {C.}~\bibnamefont {Gutgsell}}, \bibinfo {author} {\bibfnamefont
  {C.}~\bibnamefont {Nistor}}, \bibinfo {author} {\bibfnamefont {C.~L.}\
  \bibnamefont {Degen}}, \bibinfo {author} {\bibfnamefont {M.}~\bibnamefont
  {Trassin}}, \bibinfo {author} {\bibfnamefont {M.}~\bibnamefont {Fiebig}},\
  and\ \bibinfo {author} {\bibfnamefont {P.}~\bibnamefont {Gambardella}},\
  }\bibfield  {title} {\bibinfo {title} {High-speed domain wall racetracks in a
  magnetic insulator},\ }\href {https://doi.org/10.1038/s41467-019-12676-7}
  {\bibfield  {journal} {\bibinfo  {journal} {Nature Communications}\ }\textbf
  {\bibinfo {volume} {10}},\ \bibinfo {pages} {4750} (\bibinfo {year}
  {2019})}\BibitemShut {NoStop}%
\bibitem [{\citenamefont {Caretta}\ and\ \citenamefont
  {Avci}(2024)}]{caretta_apl_mat_2024}%
  \BibitemOpen
  \bibfield  {author} {\bibinfo {author} {\bibfnamefont {L.}~\bibnamefont
  {Caretta}}\ and\ \bibinfo {author} {\bibfnamefont {C.~O.}\ \bibnamefont
  {Avci}},\ }\bibfield  {title} {\bibinfo {title} {Domain walls speed up in
  insulating ferrimagnetic garnets},\ }\href
  {https://doi.org/10.1063/5.0159669} {\bibfield  {journal} {\bibinfo
  {journal} {APL Materials}\ }\textbf {\bibinfo {volume} {12}},\ \bibinfo
  {pages} {011106} (\bibinfo {year} {2024})}\BibitemShut {NoStop}%
\bibitem [{\citenamefont {Chauve}\ \emph {et~al.}(2000)\citenamefont {Chauve},
  \citenamefont {Giamarchi},\ and\ \citenamefont {Le~Doussal}}]{chauve_2000}%
  \BibitemOpen
  \bibfield  {author} {\bibinfo {author} {\bibfnamefont {P.}~\bibnamefont
  {Chauve}}, \bibinfo {author} {\bibfnamefont {T.}~\bibnamefont {Giamarchi}},\
  and\ \bibinfo {author} {\bibfnamefont {P.}~\bibnamefont {Le~Doussal}},\
  }\bibfield  {title} {\bibinfo {title} {Creep and depinning in disordered
  media},\ }\href {https://doi.org/10.1103/PhysRevB.62.6241} {\bibfield
  {journal} {\bibinfo  {journal} {Phys. Rev. B}\ }\textbf {\bibinfo {volume}
  {62}},\ \bibinfo {pages} {6241} (\bibinfo {year} {2000})}\BibitemShut
  {NoStop}%
\bibitem [{\citenamefont {Ferr\'e}\ \emph {et~al.}(2013)\citenamefont
  {Ferr\'e}, \citenamefont {Metaxas}, \citenamefont {Mougin}, \citenamefont
  {Jamet}, \citenamefont {Gorchon},\ and\ \citenamefont
  {Jeudy}}]{Ferre_CRP_2013_review}%
  \BibitemOpen
  \bibfield  {author} {\bibinfo {author} {\bibfnamefont {J.}~\bibnamefont
  {Ferr\'e}}, \bibinfo {author} {\bibfnamefont {P.~J.}\ \bibnamefont
  {Metaxas}}, \bibinfo {author} {\bibfnamefont {A.}~\bibnamefont {Mougin}},
  \bibinfo {author} {\bibfnamefont {J.-P.}\ \bibnamefont {Jamet}}, \bibinfo
  {author} {\bibfnamefont {J.}~\bibnamefont {Gorchon}},\ and\ \bibinfo {author}
  {\bibfnamefont {V.}~\bibnamefont {Jeudy}},\ }\bibfield  {title} {\bibinfo
  {title} {Universal magnetic domain wall dynamics in the presence of weak
  disorder},\ }\href
  {https://doi.org/http://dx.doi.org/10.1016/j.crhy.2013.08.001} {\bibfield
  {journal} {\bibinfo  {journal} {Comptes Rendus Physique}\ }\textbf {\bibinfo
  {volume} {14}},\ \bibinfo {pages} {651 } (\bibinfo {year}
  {2013})}\BibitemShut {NoStop}%
\bibitem [{\citenamefont {de~Leeuw}\ \emph {et~al.}(1980)\citenamefont
  {de~Leeuw}, \citenamefont {van~den Doel},\ and\ \citenamefont
  {Enz}}]{De_Leeuw_RepProgPhys_1980}%
  \BibitemOpen
  \bibfield  {author} {\bibinfo {author} {\bibfnamefont {F.~H.}\ \bibnamefont
  {de~Leeuw}}, \bibinfo {author} {\bibfnamefont {R.}~\bibnamefont {van~den
  Doel}},\ and\ \bibinfo {author} {\bibfnamefont {U.}~\bibnamefont {Enz}},\
  }\bibfield  {title} {\bibinfo {title} {Dynamic properties of magnetic domain
  walls and magnetic bubbles},\ }\href
  {https://doi.org/10.1088/0034-4885/43/6/001} {\bibfield  {journal} {\bibinfo
  {journal} {Reports on Progress in Physics}\ }\textbf {\bibinfo {volume}
  {43}},\ \bibinfo {pages} {689} (\bibinfo {year} {1980})}\BibitemShut
  {NoStop}%
\bibitem [{\citenamefont {Malozemoff}\ and\ \citenamefont
  {Slonczewski}(2016)}]{malozemoff}%
  \BibitemOpen
  \bibfield  {author} {\bibinfo {author} {\bibfnamefont {A.~P.}\ \bibnamefont
  {Malozemoff}}\ and\ \bibinfo {author} {\bibfnamefont {J.~C.}\ \bibnamefont
  {Slonczewski}},\ }\href@noop {} {\emph {\bibinfo {title} {Magnetic Domain
  Walls in Bubble Materials: Advances in Materials and Device Research}}},\
  Vol.~\bibinfo {volume} {1}\ (\bibinfo  {publisher} {Academic press},\
  \bibinfo {year} {2016})\BibitemShut {NoStop}%
\bibitem [{\citenamefont {Wei}\ \emph {et~al.}(2022)\citenamefont {Wei},
  \citenamefont {Santos}, \citenamefont {Lusero}, \citenamefont {Bauer},
  \citenamefont {Ben~Youssef},\ and\ \citenamefont {van
  Wees}}]{Wei_NatMat_2022}%
  \BibitemOpen
  \bibfield  {author} {\bibinfo {author} {\bibfnamefont {X.-Y.}\ \bibnamefont
  {Wei}}, \bibinfo {author} {\bibfnamefont {O.~A.}\ \bibnamefont {Santos}},
  \bibinfo {author} {\bibfnamefont {C.~H.~S.}\ \bibnamefont {Lusero}}, \bibinfo
  {author} {\bibfnamefont {G.~E.~W.}\ \bibnamefont {Bauer}}, \bibinfo {author}
  {\bibfnamefont {J.}~\bibnamefont {Ben~Youssef}},\ and\ \bibinfo {author}
  {\bibfnamefont {B.~J.}\ \bibnamefont {van Wees}},\ }\bibfield  {title}
  {\bibinfo {title} {Giant magnon spin conductivity in ultrathin yttrium iron
  garnet films},\ }\href {https://doi.org/10.1038/s41563-022-01369-0}
  {\bibfield  {journal} {\bibinfo  {journal} {Nature Materials}\ }\textbf
  {\bibinfo {volume} {21}},\ \bibinfo {pages} {1352} (\bibinfo {year}
  {2022})}\BibitemShut {NoStop}%
\bibitem [{\citenamefont {Jeudy}\ \emph {et~al.}(2018)\citenamefont {Jeudy},
  \citenamefont {D\'{\i}az~Pardo}, \citenamefont {Savero~Torres}, \citenamefont
  {Bustingorry},\ and\ \citenamefont {Kolton}}]{jeudy_PRB_2018_DW_pinning}%
  \BibitemOpen
  \bibfield  {author} {\bibinfo {author} {\bibfnamefont {V.}~\bibnamefont
  {Jeudy}}, \bibinfo {author} {\bibfnamefont {R.}~\bibnamefont
  {D\'{\i}az~Pardo}}, \bibinfo {author} {\bibfnamefont {W.}~\bibnamefont
  {Savero~Torres}}, \bibinfo {author} {\bibfnamefont {S.}~\bibnamefont
  {Bustingorry}},\ and\ \bibinfo {author} {\bibfnamefont {A.~B.}\ \bibnamefont
  {Kolton}},\ }\bibfield  {title} {\bibinfo {title} {Pinning of domain walls in
  thin ferromagnetic films},\ }\href
  {https://doi.org/10.1103/PhysRevB.98.054406} {\bibfield  {journal} {\bibinfo
  {journal} {Phys. Rev. B}\ }\textbf {\bibinfo {volume} {98}},\ \bibinfo
  {pages} {054406} (\bibinfo {year} {2018})}\BibitemShut {NoStop}%
\bibitem [{\citenamefont {Schryer}\ and\ \citenamefont
  {Walker}(1974)}]{Schryer_JAP_1974}%
  \BibitemOpen
  \bibfield  {author} {\bibinfo {author} {\bibfnamefont {N.~L.}\ \bibnamefont
  {Schryer}}\ and\ \bibinfo {author} {\bibfnamefont {L.~R.}\ \bibnamefont
  {Walker}},\ }\bibfield  {title} {\bibinfo {title} {The motion of 180° domain
  walls in uniform dc magnetic fields},\ }\href
  {https://doi.org/10.1063/1.1663252} {\bibfield  {journal} {\bibinfo
  {journal} {Journal of Applied Physics}\ }\textbf {\bibinfo {volume} {45}},\
  \bibinfo {pages} {5406} (\bibinfo {year} {1974})}\BibitemShut {NoStop}%
\bibitem [{\citenamefont {Thiaville}\ \emph {et~al.}(2012)\citenamefont
  {Thiaville}, \citenamefont {Rohart}, \citenamefont {Ju\'e}, \citenamefont
  {Cros},\ and\ \citenamefont {A.~Fert}}]{thiaville_EPL_2012}%
  \BibitemOpen
  \bibfield  {author} {\bibinfo {author} {\bibfnamefont {A.}~\bibnamefont
  {Thiaville}}, \bibinfo {author} {\bibfnamefont {S.}~\bibnamefont {Rohart}},
  \bibinfo {author} {\bibfnamefont {E.}~\bibnamefont {Ju\'e}}, \bibinfo
  {author} {\bibfnamefont {V.}~\bibnamefont {Cros}},\ and\ \bibinfo {author}
  {\bibfnamefont {A.}~\bibnamefont {A.~Fert}},\ }\bibfield  {title} {\bibinfo
  {title} {Dynamics of dzyaloshinskii domain walls in ultrathin magnetic
  films},\ }\href {https://doi.org/10.1209/0295-5075/100/57002} {\bibfield
  {journal} {\bibinfo  {journal} {EPL}\ }\textbf {\bibinfo {volume} {100}},\
  \bibinfo {pages} {57002} (\bibinfo {year} {2012})}\BibitemShut {NoStop}%
\bibitem [{\citenamefont {Yamada}\ \emph {et~al.}(2011)\citenamefont {Yamada},
  \citenamefont {Jamet}, \citenamefont {Nakatani}, \citenamefont {Mougin},
  \citenamefont {Thiaville}, \citenamefont {Ono},\ and\ \citenamefont
  {Ferré}}]{yamada_apex_2011}%
  \BibitemOpen
  \bibfield  {author} {\bibinfo {author} {\bibfnamefont {K.}~\bibnamefont
  {Yamada}}, \bibinfo {author} {\bibfnamefont {J.-P.}\ \bibnamefont {Jamet}},
  \bibinfo {author} {\bibfnamefont {Y.}~\bibnamefont {Nakatani}}, \bibinfo
  {author} {\bibfnamefont {A.}~\bibnamefont {Mougin}}, \bibinfo {author}
  {\bibfnamefont {A.}~\bibnamefont {Thiaville}}, \bibinfo {author}
  {\bibfnamefont {T.}~\bibnamefont {Ono}},\ and\ \bibinfo {author}
  {\bibfnamefont {J.}~\bibnamefont {Ferré}},\ }\bibfield  {title} {\bibinfo
  {title} {Influence of instabilities on high-field magnetic domain wall
  velocity in {(Co/Ni)} nanostrips},\ }\href
  {http://stacks.iop.org/1882-0786/4/i=11/a=113001} {\bibfield  {journal}
  {\bibinfo  {journal} {Applied Physics Express}\ }\textbf {\bibinfo {volume}
  {4}},\ \bibinfo {pages} {113001} (\bibinfo {year} {2011})}\BibitemShut
  {NoStop}%
\bibitem [{\citenamefont {Herranen}\ and\ \citenamefont
  {Laurson}(2019)}]{herranen_PRL_2019}%
  \BibitemOpen
  \bibfield  {author} {\bibinfo {author} {\bibfnamefont {T.}~\bibnamefont
  {Herranen}}\ and\ \bibinfo {author} {\bibfnamefont {L.}~\bibnamefont
  {Laurson}},\ }\bibfield  {title} {\bibinfo {title} {Barkhausen noise from
  precessional domain wall motion},\ }\href
  {https://doi.org/10.1103/PhysRevLett.122.117205} {\bibfield  {journal}
  {\bibinfo  {journal} {Phys. Rev. Lett.}\ }\textbf {\bibinfo {volume} {122}},\
  \bibinfo {pages} {117205} (\bibinfo {year} {2019})}\BibitemShut {NoStop}%
\bibitem [{\citenamefont {Ju\'e}\ \emph {et~al.}(2016)\citenamefont {Ju\'e},
  \citenamefont {Thiaville}, \citenamefont {Pizzini}, \citenamefont {Miltat},
  \citenamefont {Sampaio}, \citenamefont {Buda-Prejbeanu}, \citenamefont
  {Rohart}, \citenamefont {Vogel}, \citenamefont {Bonfim}, \citenamefont
  {Boulle}, \citenamefont {Auffret}, \citenamefont {Miron},\ and\ \citenamefont
  {Gaudin}}]{jue_PRB_2016}%
  \BibitemOpen
  \bibfield  {author} {\bibinfo {author} {\bibfnamefont {E.}~\bibnamefont
  {Ju\'e}}, \bibinfo {author} {\bibfnamefont {A.}~\bibnamefont {Thiaville}},
  \bibinfo {author} {\bibfnamefont {S.}~\bibnamefont {Pizzini}}, \bibinfo
  {author} {\bibfnamefont {J.}~\bibnamefont {Miltat}}, \bibinfo {author}
  {\bibfnamefont {J.}~\bibnamefont {Sampaio}}, \bibinfo {author} {\bibfnamefont
  {L.~D.}\ \bibnamefont {Buda-Prejbeanu}}, \bibinfo {author} {\bibfnamefont
  {S.}~\bibnamefont {Rohart}}, \bibinfo {author} {\bibfnamefont
  {J.}~\bibnamefont {Vogel}}, \bibinfo {author} {\bibfnamefont
  {M.}~\bibnamefont {Bonfim}}, \bibinfo {author} {\bibfnamefont
  {O.}~\bibnamefont {Boulle}}, \bibinfo {author} {\bibfnamefont
  {S.}~\bibnamefont {Auffret}}, \bibinfo {author} {\bibfnamefont {I.~M.}\
  \bibnamefont {Miron}},\ and\ \bibinfo {author} {\bibfnamefont
  {G.}~\bibnamefont {Gaudin}},\ }\bibfield  {title} {\bibinfo {title} {Domain
  wall dynamics in ultrathin {Pt/Co/AlO$_x$} microstrips under large combined
  magnetic fields},\ }\href {https://doi.org/10.1103/PhysRevB.93.014403}
  {\bibfield  {journal} {\bibinfo  {journal} {Phys. Rev. B}\ }\textbf {\bibinfo
  {volume} {93}},\ \bibinfo {pages} {014403} (\bibinfo {year}
  {2016})}\BibitemShut {NoStop}%
\bibitem [{\citenamefont {Moretti}\ \emph {et~al.}(2017)\citenamefont
  {Moretti}, \citenamefont {Voto},\ and\ \citenamefont
  {Martinez}}]{moretti_PRB_2017}%
  \BibitemOpen
  \bibfield  {author} {\bibinfo {author} {\bibfnamefont {S.}~\bibnamefont
  {Moretti}}, \bibinfo {author} {\bibfnamefont {M.}~\bibnamefont {Voto}},\ and\
  \bibinfo {author} {\bibfnamefont {E.}~\bibnamefont {Martinez}},\ }\bibfield
  {title} {\bibinfo {title} {Dynamical depinning of chiral domain walls},\
  }\href {https://doi.org/10.1103/PhysRevB.96.054433} {\bibfield  {journal}
  {\bibinfo  {journal} {Phys. Rev. B}\ }\textbf {\bibinfo {volume} {96}},\
  \bibinfo {pages} {054433} (\bibinfo {year} {2017})}\BibitemShut {NoStop}%
\bibitem [{\citenamefont {Lemerle}\ \emph {et~al.}(1998)\citenamefont
  {Lemerle}, \citenamefont {Ferr{\'e}}, \citenamefont {Chappert}, \citenamefont
  {Mathet}, \citenamefont {Giamarchi},\ and\ \citenamefont {{Le
  Doussal}}}]{lemerle_PRL_1998_domainwall_creep}%
  \BibitemOpen
  \bibfield  {author} {\bibinfo {author} {\bibfnamefont {S.}~\bibnamefont
  {Lemerle}}, \bibinfo {author} {\bibfnamefont {J.}~\bibnamefont {Ferr{\'e}}},
  \bibinfo {author} {\bibfnamefont {C.}~\bibnamefont {Chappert}}, \bibinfo
  {author} {\bibfnamefont {V.}~\bibnamefont {Mathet}}, \bibinfo {author}
  {\bibfnamefont {T.}~\bibnamefont {Giamarchi}},\ and\ \bibinfo {author}
  {\bibfnamefont {P.}~\bibnamefont {{Le Doussal}}},\ }\bibfield  {title}
  {\bibinfo {title} {Domain wall creep in an ising ultrathin magnetic film},\
  }\href@noop {} {\bibfield  {journal} {\bibinfo  {journal} {Phys. Rev. Lett.}\
  }\textbf {\bibinfo {volume} {80}},\ \bibinfo {pages} {849} (\bibinfo {year}
  {1998})}\BibitemShut {NoStop}%
\bibitem [{\citenamefont {Albornoz}\ \emph {et~al.}(2024)\citenamefont
  {Albornoz}, \citenamefont {Pardo}, \citenamefont {Lema\^{\i}tre},
  \citenamefont {Bustingorry}, \citenamefont {Curiale},\ and\ \citenamefont
  {Jeudy}}]{albornoz_PRB_2024}%
  \BibitemOpen
  \bibfield  {author} {\bibinfo {author} {\bibfnamefont {L.~J.}\ \bibnamefont
  {Albornoz}}, \bibinfo {author} {\bibfnamefont {R.~D.}\ \bibnamefont {Pardo}},
  \bibinfo {author} {\bibfnamefont {A.}~\bibnamefont {Lema\^{\i}tre}}, \bibinfo
  {author} {\bibfnamefont {S.}~\bibnamefont {Bustingorry}}, \bibinfo {author}
  {\bibfnamefont {J.}~\bibnamefont {Curiale}},\ and\ \bibinfo {author}
  {\bibfnamefont {V.}~\bibnamefont {Jeudy}},\ }\bibfield  {title} {\bibinfo
  {title} {Internal structure dependent creep motion of domain walls driven by
  spin-transfer torques},\ }\href {https://doi.org/10.1103/PhysRevB.110.024403}
  {\bibfield  {journal} {\bibinfo  {journal} {Phys. Rev. B}\ }\textbf {\bibinfo
  {volume} {110}},\ \bibinfo {pages} {024403} (\bibinfo {year}
  {2024})}\BibitemShut {NoStop}%
\bibitem [{\citenamefont {Jeudy}\ \emph {et~al.}(2016)\citenamefont {Jeudy},
  \citenamefont {Mougin}, \citenamefont {Bustingorry}, \citenamefont
  {Savero~Torres}, \citenamefont {Gorchon}, \citenamefont {Kolton},
  \citenamefont {Lema\^{\i}tre},\ and\ \citenamefont
  {Jamet}}]{jeudy_PRL_2016_energy_barrier}%
  \BibitemOpen
  \bibfield  {author} {\bibinfo {author} {\bibfnamefont {V.}~\bibnamefont
  {Jeudy}}, \bibinfo {author} {\bibfnamefont {A.}~\bibnamefont {Mougin}},
  \bibinfo {author} {\bibfnamefont {S.}~\bibnamefont {Bustingorry}}, \bibinfo
  {author} {\bibfnamefont {W.}~\bibnamefont {Savero~Torres}}, \bibinfo {author}
  {\bibfnamefont {J.}~\bibnamefont {Gorchon}}, \bibinfo {author} {\bibfnamefont
  {A.~B.}\ \bibnamefont {Kolton}}, \bibinfo {author} {\bibfnamefont
  {A.}~\bibnamefont {Lema\^{\i}tre}},\ and\ \bibinfo {author} {\bibfnamefont
  {J.-P.}\ \bibnamefont {Jamet}},\ }\bibfield  {title} {\bibinfo {title}
  {Universal pinning energy barrier for driven domain walls in thin
  ferromagnetic films},\ }\href
  {https://doi.org/10.1103/PhysRevLett.117.057201} {\bibfield  {journal}
  {\bibinfo  {journal} {Phys. Rev. Lett.}\ }\textbf {\bibinfo {volume} {117}},\
  \bibinfo {pages} {057201} (\bibinfo {year} {2016})}\BibitemShut {NoStop}%
\bibitem [{\citenamefont {Diaz~Pardo}\ \emph {et~al.}(2017)\citenamefont
  {Diaz~Pardo}, \citenamefont {Savero~Torres}, \citenamefont {Kolton},
  \citenamefont {Bustingorry},\ and\ \citenamefont
  {Jeudy}}]{diaz_PRB_2017_depinning}%
  \BibitemOpen
  \bibfield  {author} {\bibinfo {author} {\bibfnamefont {R.}~\bibnamefont
  {Diaz~Pardo}}, \bibinfo {author} {\bibfnamefont {W.}~\bibnamefont
  {Savero~Torres}}, \bibinfo {author} {\bibfnamefont {A.~B.}\ \bibnamefont
  {Kolton}}, \bibinfo {author} {\bibfnamefont {S.}~\bibnamefont
  {Bustingorry}},\ and\ \bibinfo {author} {\bibfnamefont {V.}~\bibnamefont
  {Jeudy}},\ }\bibfield  {title} {\bibinfo {title} {Universal depinning
  transition of domain walls in ultrathin ferromagnets},\ }\href
  {https://doi.org/10.1103/PhysRevB.95.184434} {\bibfield  {journal} {\bibinfo
  {journal} {Phys. Rev. B}\ }\textbf {\bibinfo {volume} {95}},\ \bibinfo
  {pages} {184434} (\bibinfo {year} {2017})}\BibitemShut {NoStop}%
\bibitem [{\citenamefont {Kolton}\ \emph {et~al.}(2009)\citenamefont {Kolton},
  \citenamefont {Rosso}, \citenamefont {Giamarchi},\ and\ \citenamefont
  {Krauth}}]{kolton_prb_2009_pathways}%
  \BibitemOpen
  \bibfield  {author} {\bibinfo {author} {\bibfnamefont {A.~B.}\ \bibnamefont
  {Kolton}}, \bibinfo {author} {\bibfnamefont {A.}~\bibnamefont {Rosso}},
  \bibinfo {author} {\bibfnamefont {T.}~\bibnamefont {Giamarchi}},\ and\
  \bibinfo {author} {\bibfnamefont {W.}~\bibnamefont {Krauth}},\ }\bibfield
  {title} {\bibinfo {title} {Creep dynamics of elastic manifolds via exact
  transition pathways},\ }\href@noop {} {\bibfield  {journal} {\bibinfo
  {journal} {Phys. Rev. B}\ }\textbf {\bibinfo {volume} {79}},\ \bibinfo
  {pages} {184207} (\bibinfo {year} {2009})}\BibitemShut {NoStop}%
\bibitem [{\citenamefont {Caballero}\ \emph {et~al.}(2018)\citenamefont
  {Caballero}, \citenamefont {Ferrero}, \citenamefont {Kolton}, \citenamefont
  {Curiale}, \citenamefont {Jeudy},\ and\ \citenamefont
  {Bustingorry}}]{caballero_PRE_2018}%
  \BibitemOpen
  \bibfield  {author} {\bibinfo {author} {\bibfnamefont {N.~B.}\ \bibnamefont
  {Caballero}}, \bibinfo {author} {\bibfnamefont {E.~E.}\ \bibnamefont
  {Ferrero}}, \bibinfo {author} {\bibfnamefont {A.~B.}\ \bibnamefont {Kolton}},
  \bibinfo {author} {\bibfnamefont {J.}~\bibnamefont {Curiale}}, \bibinfo
  {author} {\bibfnamefont {V.}~\bibnamefont {Jeudy}},\ and\ \bibinfo {author}
  {\bibfnamefont {S.}~\bibnamefont {Bustingorry}},\ }\bibfield  {title}
  {\bibinfo {title} {Magnetic domain wall creep and depinning: A scalar field
  model approach},\ }\href {https://doi.org/10.1103/PhysRevE.97.062122}
  {\bibfield  {journal} {\bibinfo  {journal} {Phys. Rev. E}\ }\textbf {\bibinfo
  {volume} {97}},\ \bibinfo {pages} {062122} (\bibinfo {year}
  {2018})}\BibitemShut {NoStop}%
\bibitem [{\citenamefont {Guruciaga}\ \emph {et~al.}(2021)\citenamefont
  {Guruciaga}, \citenamefont {Caballero}, \citenamefont {Jeudy}, \citenamefont
  {Curiale},\ and\ \citenamefont {Bustingorry}}]{guruciaga_JStatMech_2021}%
  \BibitemOpen
  \bibfield  {author} {\bibinfo {author} {\bibfnamefont {P.~C.}\ \bibnamefont
  {Guruciaga}}, \bibinfo {author} {\bibfnamefont {N.}~\bibnamefont
  {Caballero}}, \bibinfo {author} {\bibfnamefont {V.}~\bibnamefont {Jeudy}},
  \bibinfo {author} {\bibfnamefont {J.}~\bibnamefont {Curiale}},\ and\ \bibinfo
  {author} {\bibfnamefont {S.}~\bibnamefont {Bustingorry}},\ }\bibfield
  {title} {\bibinfo {title} {Tuning ginzburg–landau theory to quantitatively
  study thin ferromagnetic materials},\ }\href
  {https://doi.org/10.1088/1742-5468/abe40a} {\bibfield  {journal} {\bibinfo
  {journal} {Journal of Statistical Mechanics: Theory and Experiment}\ }\textbf
  {\bibinfo {volume} {2021}},\ \bibinfo {pages} {033211} (\bibinfo {year}
  {2021})}\BibitemShut {NoStop}%
\bibitem [{\citenamefont {Ben~Youssef}(1989)}]{benyoussef_thesis_1989}%
  \BibitemOpen
  \bibfield  {author} {\bibinfo {author} {\bibfnamefont {J.}~\bibnamefont
  {Ben~Youssef}},\ }\emph {\bibinfo {title} {Elaboration par épitaxie en phase
  liquide, caractérisation et étude physique des films minces de grenats
  ferrimagnetiques susbstitués par des ions bismuth}},\ \href
  {http://www.theses.fr/1989PA066048} {Ph.D. thesis} (\bibinfo {year} {1989}),\
  \bibinfo {note} {université Paris VI}\BibitemShut {NoStop}%
\bibitem [{\citenamefont {Soumah}\ \emph {et~al.}(2018)\citenamefont {Soumah},
  \citenamefont {Beaulieu}, \citenamefont {Qassym}, \citenamefont
  {Carr{\'e}t{\'e}ro}, \citenamefont {Jacquet}, \citenamefont {Lebourgeois},
  \citenamefont {Ben~Youssef}, \citenamefont {Bortolotti}, \citenamefont
  {Cros},\ and\ \citenamefont {Anane}}]{Soumah_natcom_2018}%
  \BibitemOpen
  \bibfield  {author} {\bibinfo {author} {\bibfnamefont {L.}~\bibnamefont
  {Soumah}}, \bibinfo {author} {\bibfnamefont {N.}~\bibnamefont {Beaulieu}},
  \bibinfo {author} {\bibfnamefont {L.}~\bibnamefont {Qassym}}, \bibinfo
  {author} {\bibfnamefont {C.}~\bibnamefont {Carr{\'e}t{\'e}ro}}, \bibinfo
  {author} {\bibfnamefont {E.}~\bibnamefont {Jacquet}}, \bibinfo {author}
  {\bibfnamefont {R.}~\bibnamefont {Lebourgeois}}, \bibinfo {author}
  {\bibfnamefont {J.}~\bibnamefont {Ben~Youssef}}, \bibinfo {author}
  {\bibfnamefont {P.}~\bibnamefont {Bortolotti}}, \bibinfo {author}
  {\bibfnamefont {V.}~\bibnamefont {Cros}},\ and\ \bibinfo {author}
  {\bibfnamefont {A.}~\bibnamefont {Anane}},\ }\bibfield  {title} {\bibinfo
  {title} {Ultra-low damping insulating magnetic thin films get
  perpendicular},\ }\href {https://doi.org/10.1038/s41467-018-05732-1}
  {\bibfield  {journal} {\bibinfo  {journal} {Nature Communications}\ }\textbf
  {\bibinfo {volume} {9}},\ \bibinfo {pages} {3355} (\bibinfo {year}
  {2018})}\BibitemShut {NoStop}%
\bibitem [{\citenamefont {Husain}\ \emph {et~al.}(2024)\citenamefont {Husain},
  \citenamefont {Prestes}, \citenamefont {Fayet}, \citenamefont {Collin},
  \citenamefont {Godel}, \citenamefont {Jacquet}, \citenamefont {Denneulin},
  \citenamefont {Dunin-Borkowski}, \citenamefont {Thiaville}, \citenamefont
  {Bibes}, \citenamefont {Jaffrès}, \citenamefont {Reyren}, \citenamefont
  {Fert},\ and\ \citenamefont {George}}]{husain_nanolett_2024}%
  \BibitemOpen
  \bibfield  {author} {\bibinfo {author} {\bibfnamefont {S.}~\bibnamefont
  {Husain}}, \bibinfo {author} {\bibfnamefont {N.~F.}\ \bibnamefont {Prestes}},
  \bibinfo {author} {\bibfnamefont {O.}~\bibnamefont {Fayet}}, \bibinfo
  {author} {\bibfnamefont {S.}~\bibnamefont {Collin}}, \bibinfo {author}
  {\bibfnamefont {F.}~\bibnamefont {Godel}}, \bibinfo {author} {\bibfnamefont
  {E.}~\bibnamefont {Jacquet}}, \bibinfo {author} {\bibfnamefont
  {T.}~\bibnamefont {Denneulin}}, \bibinfo {author} {\bibfnamefont {R.~E.}\
  \bibnamefont {Dunin-Borkowski}}, \bibinfo {author} {\bibfnamefont
  {A.}~\bibnamefont {Thiaville}}, \bibinfo {author} {\bibfnamefont
  {M.}~\bibnamefont {Bibes}}, \bibinfo {author} {\bibfnamefont
  {H.}~\bibnamefont {Jaffrès}}, \bibinfo {author} {\bibfnamefont
  {N.}~\bibnamefont {Reyren}}, \bibinfo {author} {\bibfnamefont
  {A.}~\bibnamefont {Fert}},\ and\ \bibinfo {author} {\bibfnamefont {J.-M.}\
  \bibnamefont {George}},\ }\bibfield  {title} {\bibinfo {title} {Field-free
  switching of perpendicular magnetization in an ultrathin epitaxial magnetic
  insulator},\ }\href {https://doi.org/10.1021/acs.nanolett.3c04413} {\bibfield
   {journal} {\bibinfo  {journal} {Nano Letters}\ }\textbf {\bibinfo {volume}
  {24}},\ \bibinfo {pages} {2743} (\bibinfo {year} {2024})}\BibitemShut
  {NoStop}%
\bibitem [{\citenamefont {Vansteenkiste}\ \emph {et~al.}(2014)\citenamefont
  {Vansteenkiste}, \citenamefont {Leliaert}, \citenamefont {Dvornik},
  \citenamefont {Helsen}, \citenamefont {Garcia-Sanchez},\ and\ \citenamefont
  {Van~Waeyenberge}}]{vansteenkiste_AIP_2014}%
  \BibitemOpen
  \bibfield  {author} {\bibinfo {author} {\bibfnamefont {A.}~\bibnamefont
  {Vansteenkiste}}, \bibinfo {author} {\bibfnamefont {J.}~\bibnamefont
  {Leliaert}}, \bibinfo {author} {\bibfnamefont {M.}~\bibnamefont {Dvornik}},
  \bibinfo {author} {\bibfnamefont {M.}~\bibnamefont {Helsen}}, \bibinfo
  {author} {\bibfnamefont {F.}~\bibnamefont {Garcia-Sanchez}},\ and\ \bibinfo
  {author} {\bibfnamefont {B.}~\bibnamefont {Van~Waeyenberge}},\ }\bibfield
  {title} {\bibinfo {title} {{The design and verification of MuMax3}},\
  }\href@noop {} {\bibfield  {journal} {\bibinfo  {journal} {AIP Advances}\
  }\textbf {\bibinfo {volume} {4}},\ \bibinfo {pages} {107133} (\bibinfo {year}
  {2014})}\BibitemShut {NoStop}%
\bibitem [{\citenamefont {Van~de Wiele}\ \emph {et~al.}(2012)\citenamefont
  {Van~de Wiele}, \citenamefont {Laurson},\ and\ \citenamefont
  {Durin}}]{van_de_wiele_PRB_2012}%
  \BibitemOpen
  \bibfield  {author} {\bibinfo {author} {\bibfnamefont {B.}~\bibnamefont
  {Van~de Wiele}}, \bibinfo {author} {\bibfnamefont {L.}~\bibnamefont
  {Laurson}},\ and\ \bibinfo {author} {\bibfnamefont {G.}~\bibnamefont
  {Durin}},\ }\bibfield  {title} {\bibinfo {title} {Effect of disorder on
  transverse domain wall dynamics in magnetic nanostrips},\ }\href
  {https://doi.org/10.1103/PhysRevB.86.144415} {\bibfield  {journal} {\bibinfo
  {journal} {Phys. Rev. B}\ }\textbf {\bibinfo {volume} {86}},\ \bibinfo
  {pages} {144415} (\bibinfo {year} {2012})}\BibitemShut {NoStop}%
\bibitem [{\citenamefont {G\'ehanne}\ \emph {et~al.}(2020)\citenamefont
  {G\'ehanne}, \citenamefont {Rohart}, \citenamefont {Thiaville},\ and\
  \citenamefont {Jeudy}}]{gehanne_PRR_2020}%
  \BibitemOpen
  \bibfield  {author} {\bibinfo {author} {\bibfnamefont {P.}~\bibnamefont
  {G\'ehanne}}, \bibinfo {author} {\bibfnamefont {S.}~\bibnamefont {Rohart}},
  \bibinfo {author} {\bibfnamefont {A.}~\bibnamefont {Thiaville}},\ and\
  \bibinfo {author} {\bibfnamefont {V.}~\bibnamefont {Jeudy}},\ }\bibfield
  {title} {\bibinfo {title} {Strength and length scale of the interaction
  between domain walls and pinning disorder in thin ferromagnetic films},\
  }\href {https://doi.org/10.1103/PhysRevResearch.2.043134} {\bibfield
  {journal} {\bibinfo  {journal} {Phys. Rev. Res.}\ }\textbf {\bibinfo {volume}
  {2}},\ \bibinfo {pages} {043134} (\bibinfo {year} {2020})}\BibitemShut
  {NoStop}%
\bibitem [{\citenamefont {Nattermann}\ \emph {et~al.}(1990)\citenamefont
  {Nattermann}, \citenamefont {Shapir},\ and\ \citenamefont
  {Vilfan}}]{nattermann_prb_1990}%
  \BibitemOpen
  \bibfield  {author} {\bibinfo {author} {\bibfnamefont {T.}~\bibnamefont
  {Nattermann}}, \bibinfo {author} {\bibfnamefont {Y.}~\bibnamefont {Shapir}},\
  and\ \bibinfo {author} {\bibfnamefont {I.}~\bibnamefont {Vilfan}},\
  }\bibfield  {title} {\bibinfo {title} {Interface pinning and dynamics in
  random systems},\ }\href {https://doi.org/10.1103/PhysRevB.42.8577}
  {\bibfield  {journal} {\bibinfo  {journal} {Phys. Rev. B}\ }\textbf {\bibinfo
  {volume} {42}},\ \bibinfo {pages} {8577} (\bibinfo {year}
  {1990})}\BibitemShut {NoStop}%
\bibitem [{\citenamefont {Dourlat}\ \emph {et~al.}(2008)\citenamefont
  {Dourlat}, \citenamefont {Jeudy}, \citenamefont {Lema\^{i}tre},\ and\
  \citenamefont {Gourdon}}]{dourlat_prb_2008}%
  \BibitemOpen
  \bibfield  {author} {\bibinfo {author} {\bibfnamefont {A.}~\bibnamefont
  {Dourlat}}, \bibinfo {author} {\bibfnamefont {V.}~\bibnamefont {Jeudy}},
  \bibinfo {author} {\bibfnamefont {A.}~\bibnamefont {Lema\^{i}tre}},\ and\
  \bibinfo {author} {\bibfnamefont {C.}~\bibnamefont {Gourdon}},\ }\bibfield
  {title} {\bibinfo {title} {Field-driven domain-wall dynamics in {(Ga,Mn)As}
  films with perpendicular anisotropy},\ }\href
  {https://doi.org/10.1103/PhysRevB.78.161303} {\bibfield  {journal} {\bibinfo
  {journal} {Phys. Rev. B}\ }\textbf {\bibinfo {volume} {78}},\ \bibinfo
  {pages} {161303} (\bibinfo {year} {2008})}\BibitemShut {NoStop}%
\bibitem [{\citenamefont {Weindler}\ \emph {et~al.}(2014)\citenamefont
  {Weindler}, \citenamefont {Bauer}, \citenamefont {Islinger}, \citenamefont
  {Boehm}, \citenamefont {Chauleau},\ and\ \citenamefont
  {Back}}]{chauleau_PRL_2014}%
  \BibitemOpen
  \bibfield  {author} {\bibinfo {author} {\bibfnamefont {T.}~\bibnamefont
  {Weindler}}, \bibinfo {author} {\bibfnamefont {H.}~\bibnamefont {Bauer}},
  \bibinfo {author} {\bibfnamefont {R.}~\bibnamefont {Islinger}}, \bibinfo
  {author} {\bibfnamefont {B.}~\bibnamefont {Boehm}}, \bibinfo {author}
  {\bibfnamefont {J.-Y.}\ \bibnamefont {Chauleau}},\ and\ \bibinfo {author}
  {\bibfnamefont {C.}~\bibnamefont {Back}},\ }\bibfield  {title} {\bibinfo
  {title} {Magnetic damping: Domain wall dynamics versus local ferromagnetic
  resonance},\ }\href
  {https://doi.org/http://dx.doi.org/10.1103/PhysRevLett.113.237204} {\bibfield
   {journal} {\bibinfo  {journal} {Phys. Rev. Lett.}\ }\textbf {\bibinfo
  {volume} {113}},\ \bibinfo {pages} {237204} (\bibinfo {year}
  {2014})}\BibitemShut {NoStop}%
\bibitem [{\citenamefont {Ferrero}\ \emph {et~al.}(2013)\citenamefont
  {Ferrero}, \citenamefont {Bustingorry},\ and\ \citenamefont
  {Kolton}}]{ferrero_PRE_2013}%
  \BibitemOpen
  \bibfield  {author} {\bibinfo {author} {\bibfnamefont {E.~E.}\ \bibnamefont
  {Ferrero}}, \bibinfo {author} {\bibfnamefont {S.}~\bibnamefont
  {Bustingorry}},\ and\ \bibinfo {author} {\bibfnamefont {A.~B.}\ \bibnamefont
  {Kolton}},\ }\bibfield  {title} {\bibinfo {title} {Nonsteady relaxation and
  critical exponents at the depinning transition},\ }\href
  {https://doi.org/10.1103/PhysRevE.87.032122} {\bibfield  {journal} {\bibinfo
  {journal} {Phys. Rev. E}\ }\textbf {\bibinfo {volume} {87}},\ \bibinfo
  {pages} {032122} (\bibinfo {year} {2013})}\BibitemShut {NoStop}%
\end{thebibliography}%
%\bibliography{
	%C:/Users/jeudy/Documents/_Publications/_En cours/Références/refs_simul
	%C:/Users/jeudy/Dropbox/manuscript_simultaneousJandH/bib/refs_creep_DMI,
	%C:/Users/jeudy/Dropbox/manuscript_simultaneousJandH/bib/paredom,
	%C:/Users/jeudy/Dropbox/manuscript_simultaneousJandH/bib/tfinita5,
	%C:/Users/jeudy/Dropbox/manuscript_simultaneousJandH/bib/dmi,	
%	}

\end{document}